\documentclass[%
 reprint,
 amsmath,amssymb,
 aps,prl
]{revtex4-1}

\usepackage{graphicx}
\usepackage{dcolumn}
\usepackage{bm}

\usepackage[caption=false]{subfig}	
\usepackage{here}

\begin{document}

\preprint{APS/123-QED}

\title{Strong Ionization in carbon Nanowires}

\author{Vural Kaymak}
\email{vural.kaymak@tp1.uni-duesseldorf.de}
\author{Alexander Pukhov}%
\affiliation{%
 Institut f\"ur Theoretische Physik, Heinrich-Heine-Universit\"at D\"usseldorf, 40225 D\"usseldorf, Germany
}%
\author{Vyacheslav N. Shlyaptsev}%
\affiliation{Department of Electrical Computer Engineering, Colorado State University, Fort Collins, Colorado 80523, USA}

\author{Jorge J. Rocca}%
\affiliation{Department of Electrical Computer Engineering, Colorado State University, Fort Collins, Colorado 80523, USA\\Department of Physics, Colorado State University, Fort Collins, Colorado 80513, USA}

\date{\today}

\begin{abstract}

Surfaces covered with nanostructures, such as nanowire arrays, have shown to facilitate a significantly higher absorption of laser energy as compared to flat surfaces. Due to the efficient coupling of the laser energy, highly energetic electrons are produced, which in turn can emit intense ultrafast X-ray pulses. In the present work we use full three dimensional PIC simulations to analyze the behavior of arrays of carbon nanowires $400 nm$ in diameter, irradiated by a $\lambda_0 = 400 nm$ laser pulse of $60 fs$ duration at FWHM and a vector potential of $a_0 = 18$. We analyze the ionization dynamics of the nanowires. We investigate the difference of the ionization strength and structure between linearly and circularly polarized laser beam. The nanowires are found to be fully ionized after about 30 laser cycles. Circularly polarized light reveals a slightly stronger ionization effect.

\begin{description}
\item[PACS numbers]
52.38.Dx, 52.38.Kd, 52.38.Ph
\end{description}
\end{abstract}

\pacs{Valid PACS appear here}
\maketitle


\section{\label{sec:intro}Introduction}

High-intensity lasers can heat solid density plasmas to high temperatures. The formation of a highly conductive plasma surface will, however, limit the penetration depth to a thin surface layer. Most of the laser light therefore is reflected. There are two different approaches to overcome this obstacle: generating a preplasma (by using a prepulse) or structuring the irradiated surface \cite{Nishikawa}.

Arrays of nanoparticles such as nanowires have been shown to facilitate a high absorption of the laser light. They typically have a low average density but a high (near-solid) local density. The high coupling of the laser energy to the material can be attributed to locally enhanced electric fields in the vicinity of the nanoparticles. Two mechanisms cause the enhancement: the so-called "lighting rod" effect, which is a purely geometric factor depending on the shape of the particle, and surface plasmon resonances \cite{Rajeev}.

Due to the high absorption of the laser energy, highly energetic electrons are produced, which can emit X-ray pulses \cite{Purvis} of down to subpicosecond duration \cite{Dorchies} of up to several hundred keV \cite{Mondal}. Designed as tabletop pulsed X-ray sources they can allow to follow processes on the atomic and molecular timescale \cite{Gibbon}. Other fields of application of laser-produced plasmas are accelerated MeV ions \cite{Zigler} and the generation of fusion neutrons \cite{Ditmire}.

\section{\label{sec:simulation_setup}Simulation Setup}

In the present work we use full three dimensional PIC simulations with the Virtual Laser Plasma Lab (VLPL) Code \cite{Pukhov}. The simulation setup contains a carbon nanowire of $400nm$ diameter and $5\mu m$ length. Their periodicitiy is $1\mu m$, which corresponds to an average density of the wires of $13\%$ solid density. A laser pulse of $60 fs$ FWHM duration and a vector potential of $a_0=18$ is injected into the simulation box and irradiates the nanowire.

\begin{figure}[H]
\centering
\includegraphics[scale=.26, angle=0]{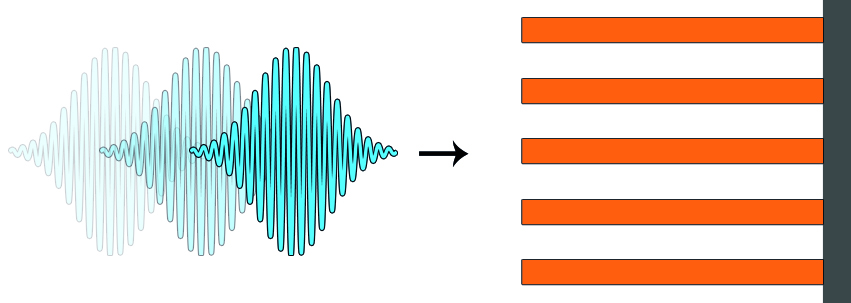}
\caption{Sketch of the simulation setup. The simulation box contains only a single rod, periodic boundary conditions in transverse direction simulate an array of nanowires.}
\label{a:setup} 
\end{figure}

The laser beam is modeled as a plane wave with a gaussian temporal profile impinging at normal incidence, see Fig. \ref{a:setup}. The code incorporates field and collisional ionization as well as binary collisions.

\section{\label{sec:ion_process}Ionization process}

We will first consider how the laser pulse penetrates into the nanowire and ionizes it along the wire axis. To point out how differently a linearly and a circularly polarized laser pulse influences the process, we subsequently consider the ionization in the plane perpendicular to the wire axis. The electrons which are removed from the carbon ions this way turn out to be pulled out of the nanowire, moving under the influence of the electric field. 

\subsection{\label{sec:ion_wire_axis}Ionization structure along the wire axis}
\begin{figure}[H]
\centering
\subfloat[\label{a:n0_lp_10}]{%
  \includegraphics[scale=0.41]{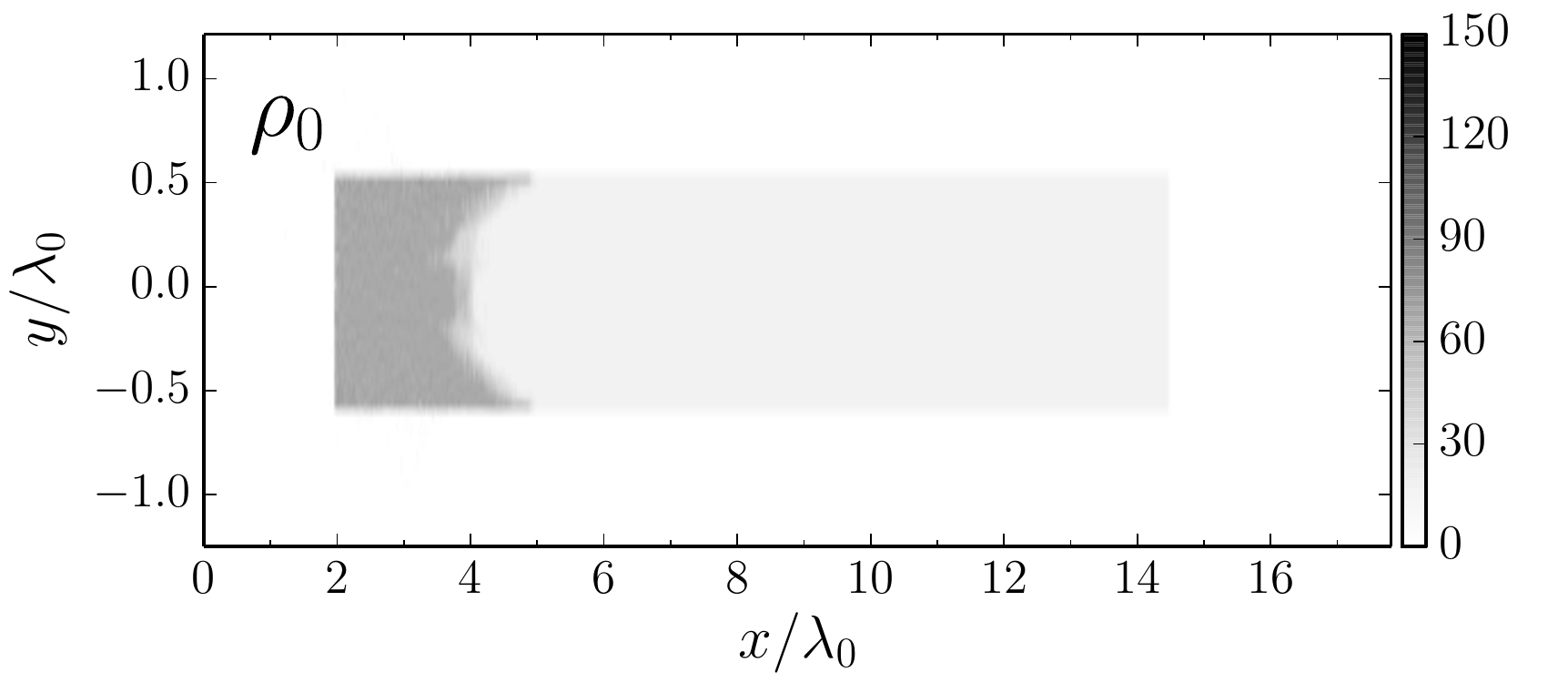} 
}\\
\subfloat[\label{a:n0_lp_20}]{%
  \includegraphics[scale=0.41]{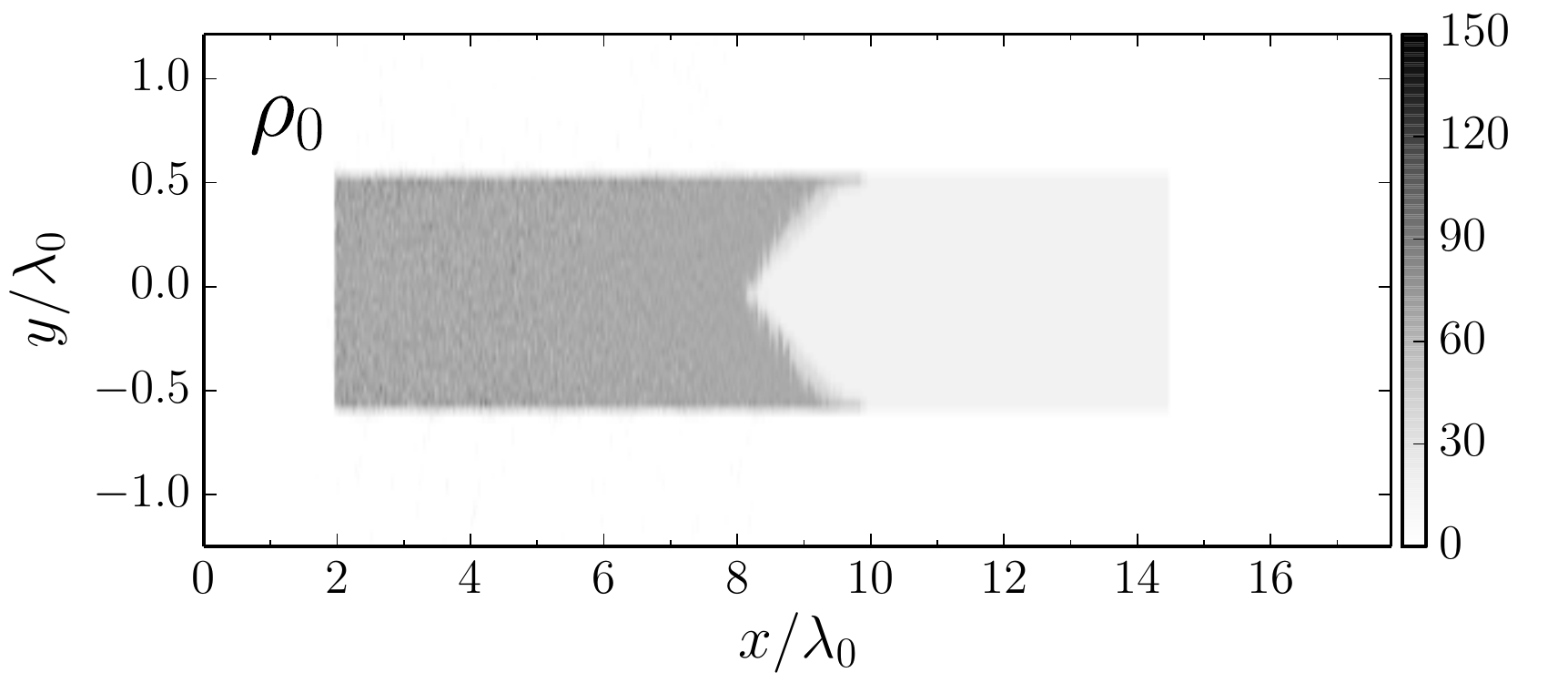} 
}\\
\subfloat[\label{a:n0_lp_30}]{%
  \includegraphics[scale=0.41]{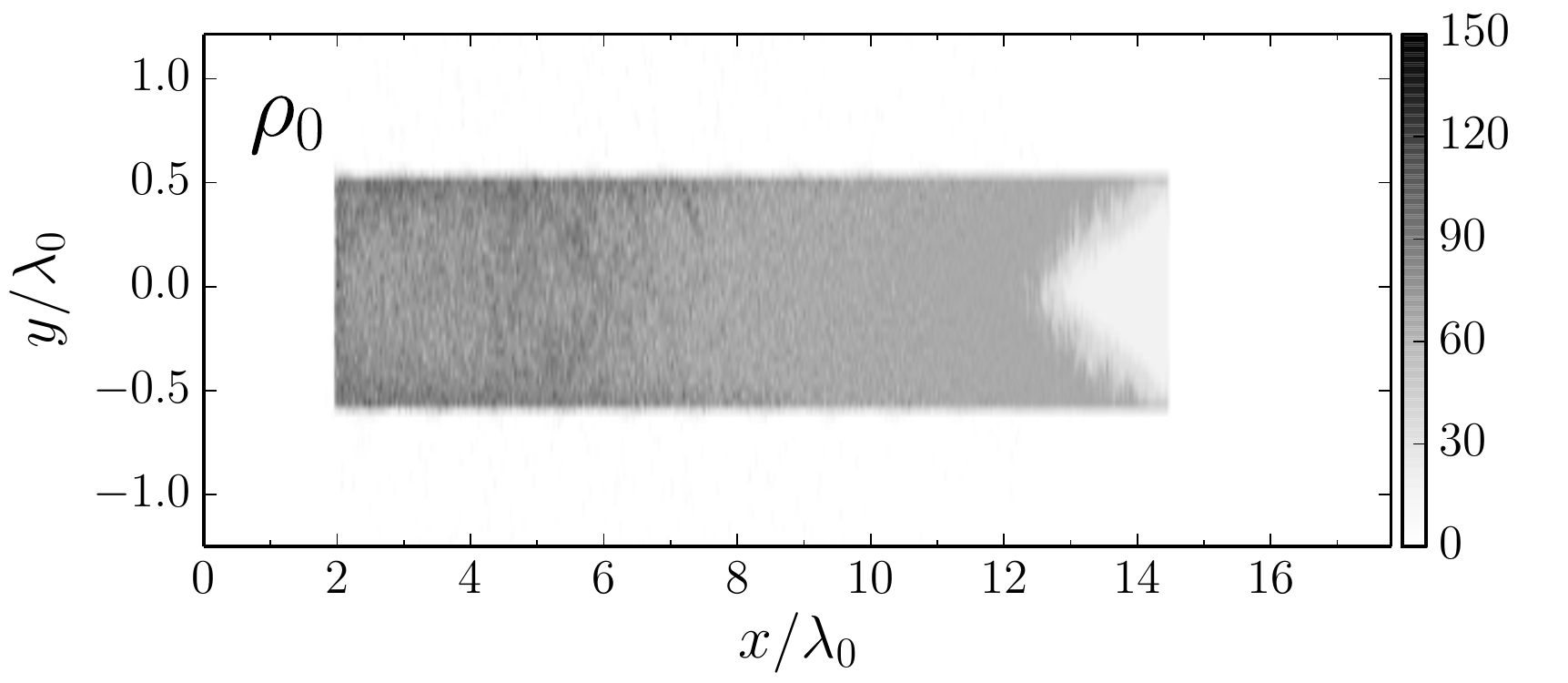} 
}\\
\subfloat[\label{a:n0_lp_40}]{%
  \includegraphics[scale=0.41]{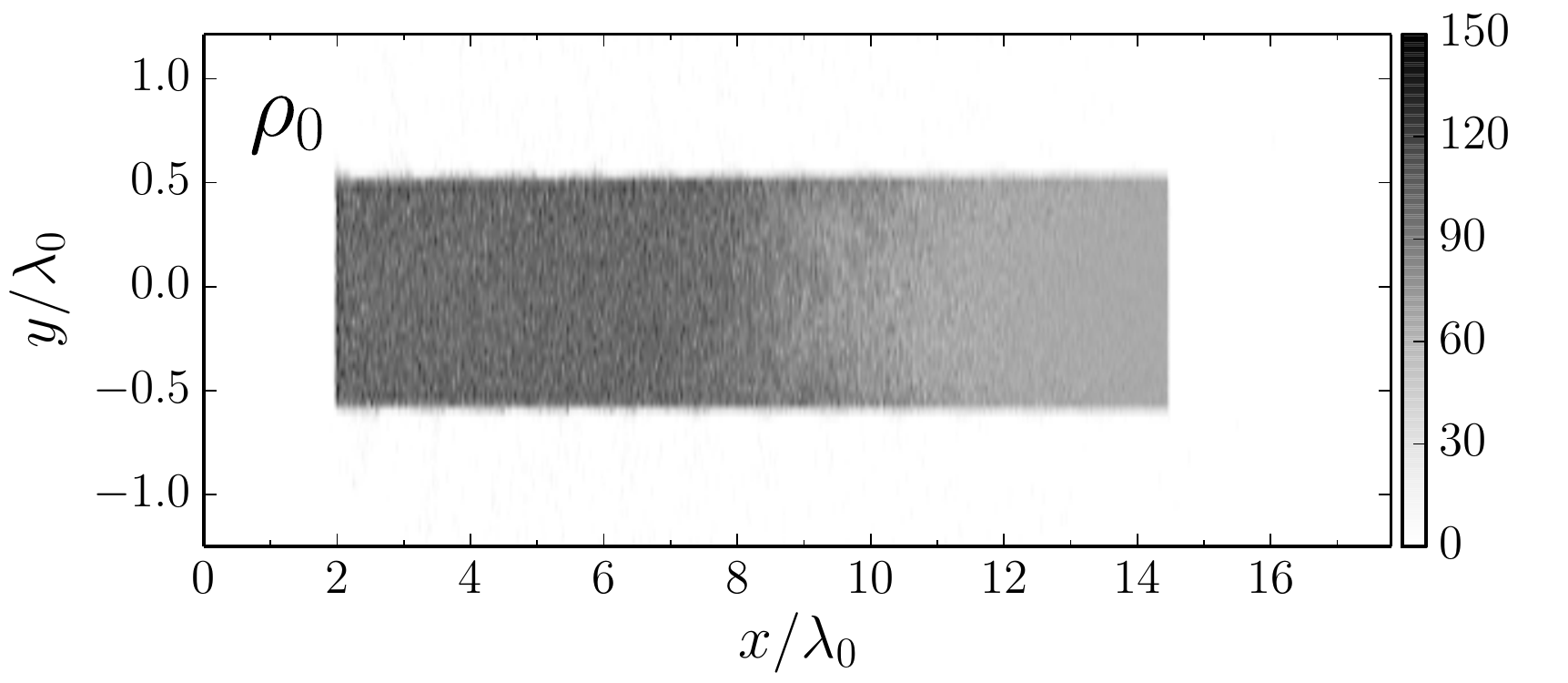} 
}
\caption{Longitudinal (cut along $z=0$) cross-sections of the electron charge density $\rho_0$ [$e n_{cr}$] ($e$: electron charge, $n_{cr}$: critical plasma density) at (a) $t=-51 T_0$, (b) $t=-46 T_0$, (c) $t=-41 T_0$ and (d) $t=-36 T_0$ before the peak of the (linearly polarized) laser pulse (in $y$-direction) hits the surface.}
\label{a:n0_lp}
\end{figure}

Figures \ref{a:n0_lp_10} - \ref{a:n0_lp_40} show the time evolution of the longitudinal cross-section (in the $y$-$x$ plane) of the electron charge density $\rho_0$.

\begin{figure}[H]
\centering
\subfloat[\label{a:Z_x_lp}]{%
  \includegraphics[scale=0.45]{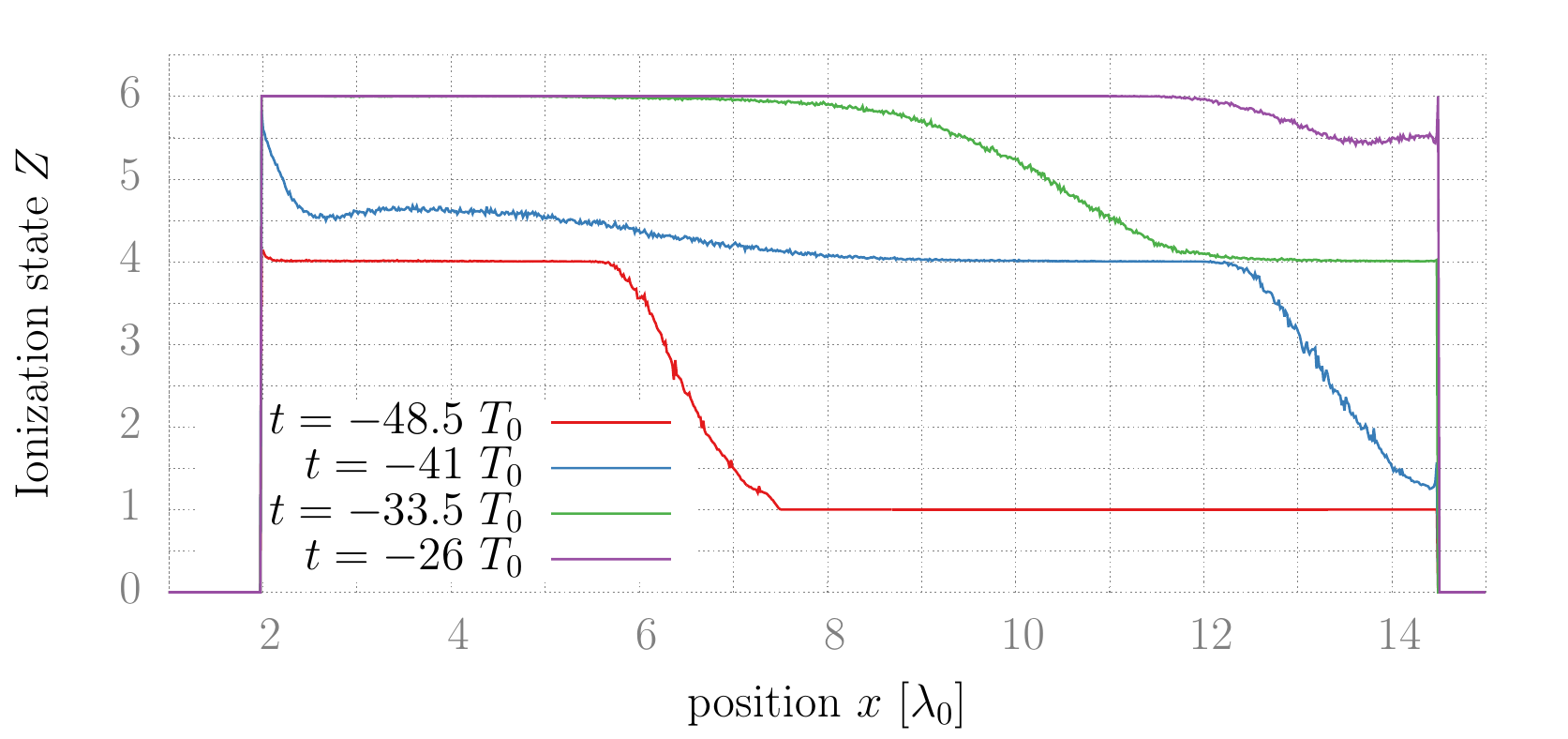} 
}\\
\subfloat[\label{a:Z_x_cp}]{%
  \includegraphics[scale=0.45]{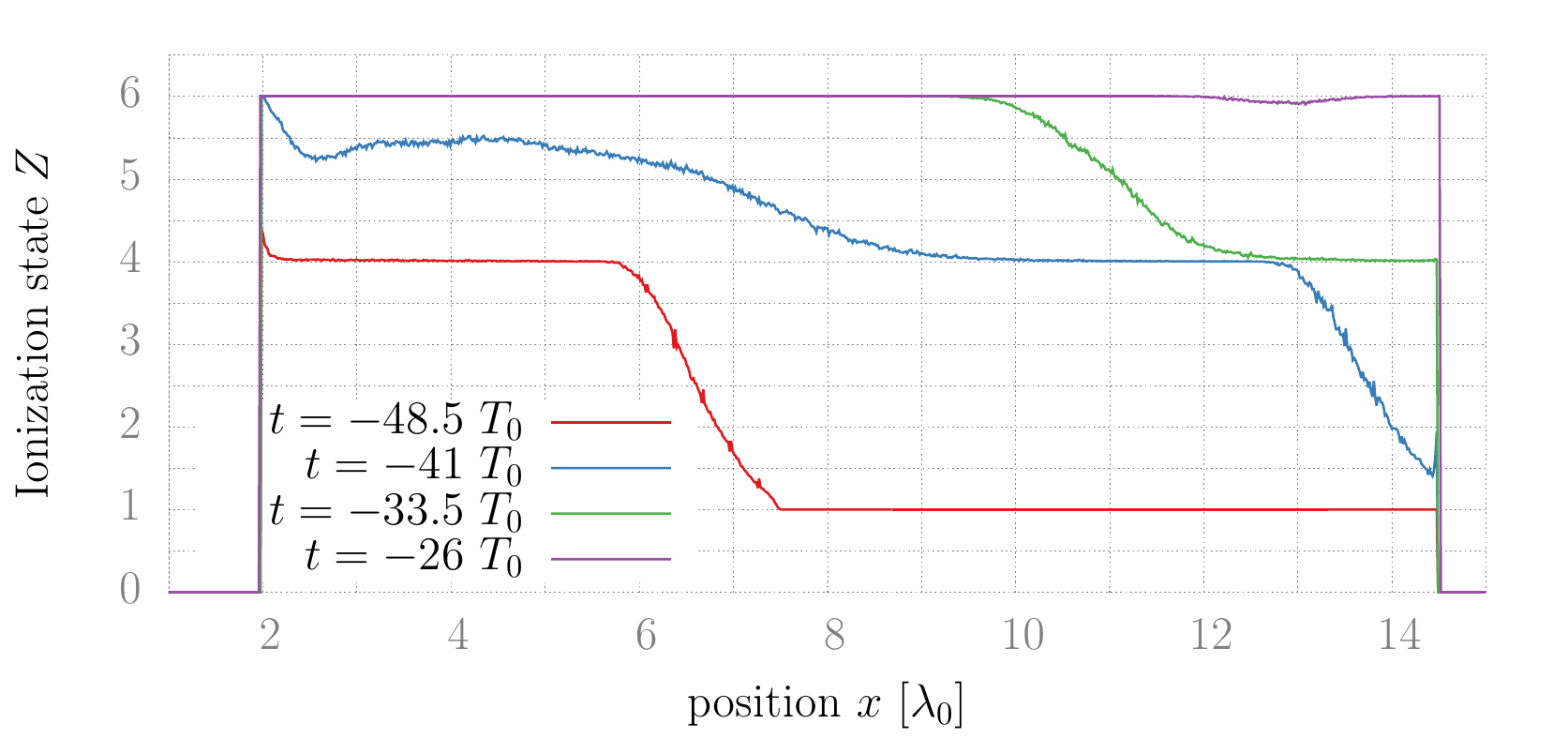} 
}
\caption{Averaged ionization state $Z$ of the carbon ions along the wire axis $x$ for a (a) linearly polarized and (b) circularly polarized laser pulse at four different times.}
\label{a:Z_x}
\end{figure}

The laser pulse, entering the simulation box from the left boundary at $x = 0$, clearly penetrates the nanowire. The carbon ions initially have an ionization state of $Z = 1$, leading to a charge density of about $\rho_0 \approx 16 e n_{cr}$, seen as the bright gray area. The tip of the wire starts to get ionized in Fig. \ref{a:n0_lp_10} and the charge density increases to about $\rho_0\approx 60 e n_{cr}$. It can be seen that the outer layer of the wire is ionized first, see the right end of the ionized area in Figures \ref{a:n0_lp_10} - \ref{a:n0_lp_30}. Only after that, since there is also collisional ionization, the ionization reaches the inner core at the wire axis. Even before the whole wire has gained a minimum charge density of about $\rho_0\approx 60 e n_{cr}$, the left half of the wire already gets ionized to values of about $\rho_0\approx 90 e n_{cr}$ (Fig. \ref{a:n0_lp_30} and \ref{a:n0_lp_40}). For figures \ref{a:n0_lp_10} - \ref{a:n0_lp_40} a linearly polarized laser was used. In principle these cross-sections do not change for circular polarization. However, this does not preclude that the ionization structure may be different. A more direct way to observe the strength of the ionization is to consider the (averaged) ionization state $Z$ of the carbon ions along the whole nanowire, see Fig. \ref{a:Z_x}. They provide a direct comparison between the ionization by a linearly polarized laser beam (Fig. \ref{a:Z_x_lp}) and a circularly polarized laser beam (Fig. \ref{a:Z_x_cp}) at four different times.

At the first timestep $t=- 48.5 T_0$ the curves for linear and circular polarization mostly coincide. The carbon wire is fourfold ionized up to $3.5 \lambda_0$ into the tip and falls to the initial state of $Z=1$ within a length of $2 \lambda_0$. $7.5$ laser cycles later, the state $Z = 4$ reaches up to $10$ wavelengths into the wire for linear polarization and $10.7$ wavelengths for circular polarization. Whereas the interval $2.5 \lambda_0 < x < 9 \lambda_0$ has averaged charged states up to $Z=4.6$ for linear polarization, the circular polarization reaches higher states of up to $Z = 5.5$. At $t = -33.5 T_0$ the first $5 \lambda_0$ laser wavelengths of the wire (linear polarization) and $7.5\lambda_0$ (circular polarization) are fully ionized. At the last step $t=-26 T_0$ the wire is basically fully ionized by the circular polarized beam. In the case of linear polarization the last $2.5$ laser wavelengths have charge states between 6 and 5.5.

\subsection{\label{sec:ion_transverse}Ionization structure in the transverse plane}

\begin{figure}[H]
\centering
\includegraphics[scale=0.4]{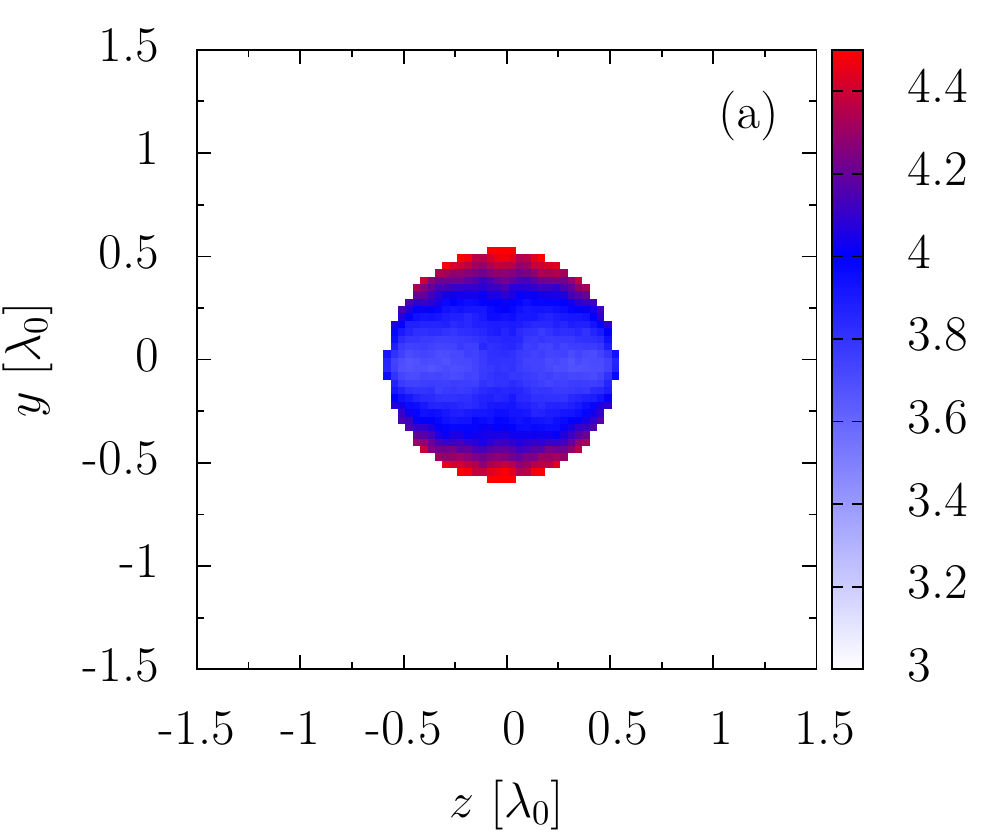}
\includegraphics[scale=0.4]{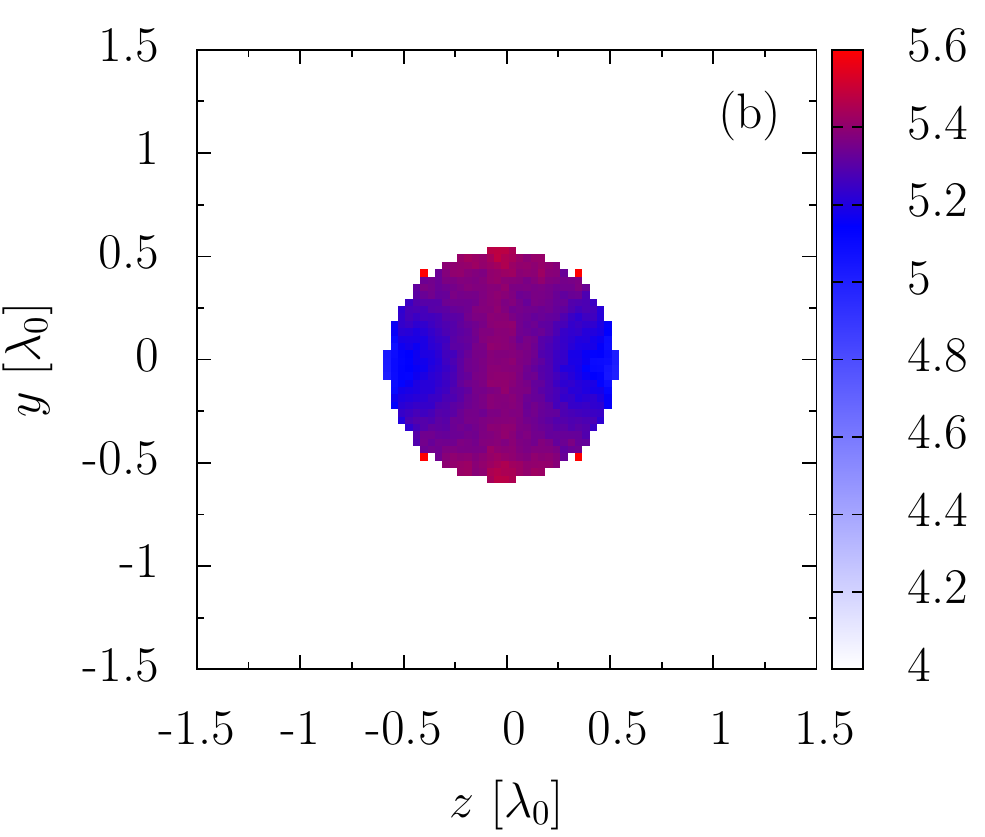}
\includegraphics[scale=0.4]{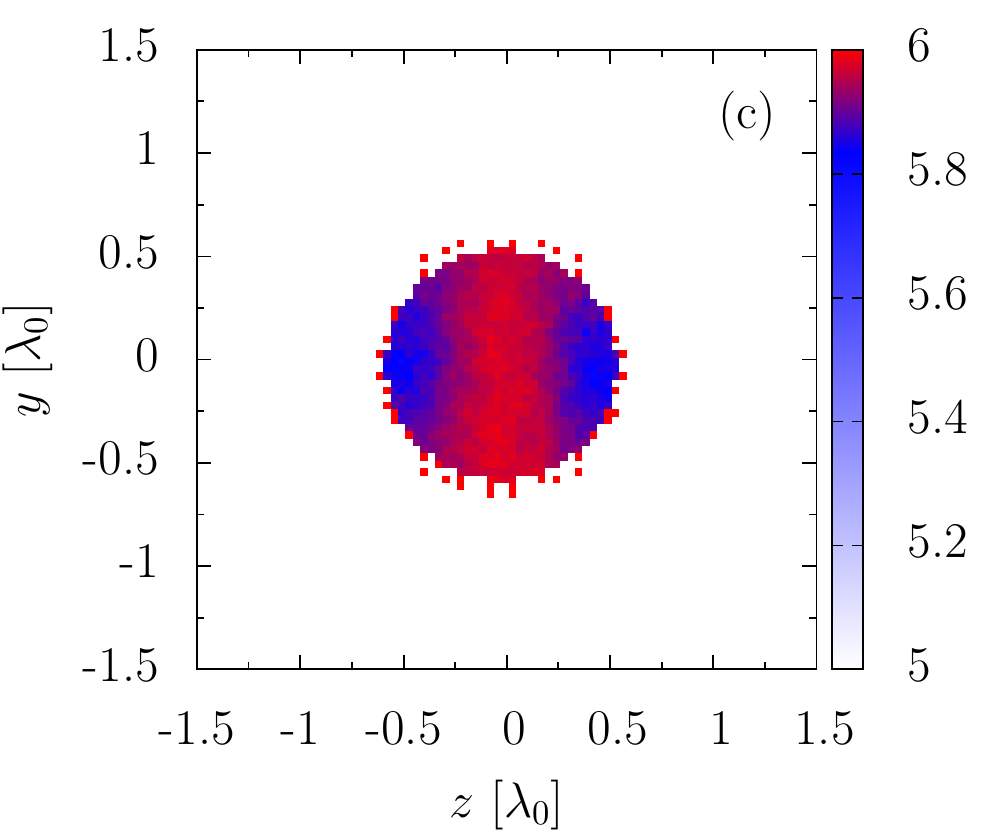}
\includegraphics[scale=0.4]{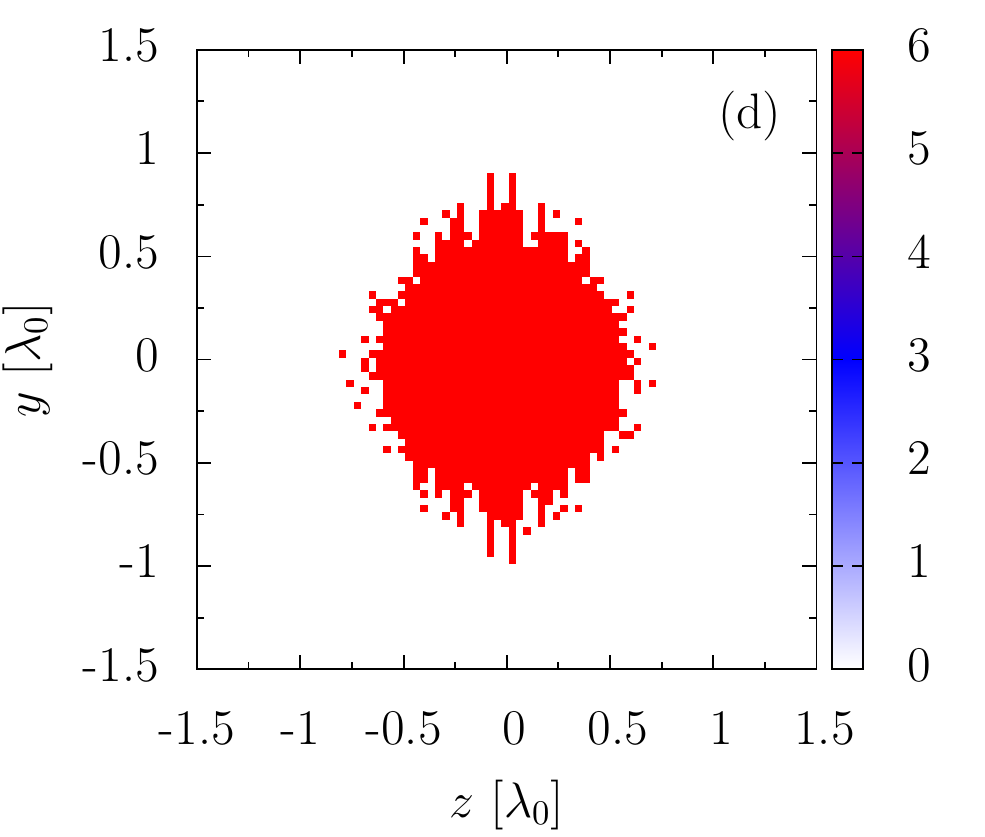}
\caption{Averaged ionization state $Z$ of the carbon ions at (a) $t = -41\ T_0$, (b) $t = -33.5\ T_0$, (c) $t = -26\ T_0$ and (d) $t = -18.5\ T_0$ before the peak of the (linearly polarized) pulse hits the surface.}
\label{a:Z_transverse_lp}
\end{figure}

A look at the transverse plane of the wire axis gives a further insight into the structure of the ionization. Figures \ref{a:Z_transverse_lp}a - \ref{a:Z_transverse_lp}d show the averaged (along the $x$ axis) ionization state in the $y$-$z$ plane for linear polarization (in $y$-direction). The white area surrounding the wire stands for empty space. In each of the figures the color scale has been adapted in order to make the difference of the charged states visible. Fig. \ref{a:Z_transverse_lp}a shows how the surfaces of the wire intersecting with the polarization direction going through the wire center carry higher ionization states (of about 4.4) than the rest of the rod. This is strong evidence that the surfaces crossing the electric field vector are ionized first. In the next subsection we will consider the electrons released this way in more detail. In the next step, the ionizing effect goes deeper into the bulk of the wire (Fig. \ref{a:Z_transverse_lp}b). Due to the linear polarization, the area of high charge states ($\approx 5.4$) goes along the $y$-direction through $x=0$. Gradually a fully ionized plasma forms in that area (\ref{a:Z_transverse_lp}c) until it covers the full rod and starts to expand (\ref{a:Z_transverse_lp}d).

\begin{figure}[H]
\centering
\includegraphics[scale=0.4]{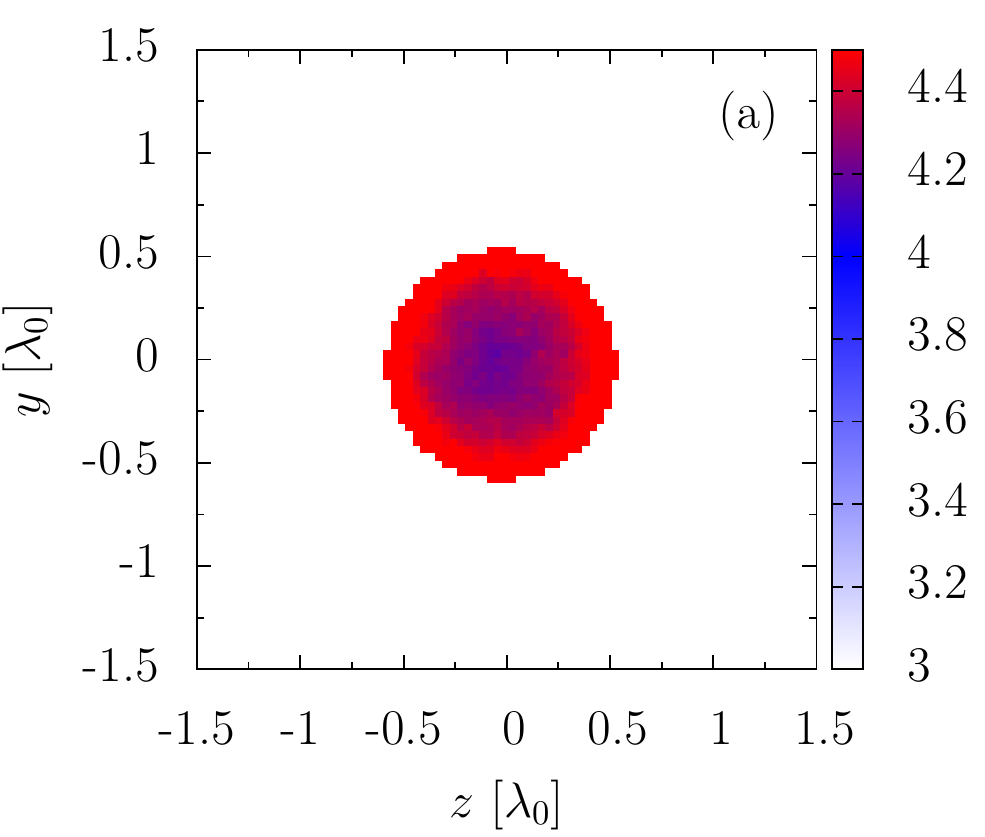}
\includegraphics[scale=0.4]{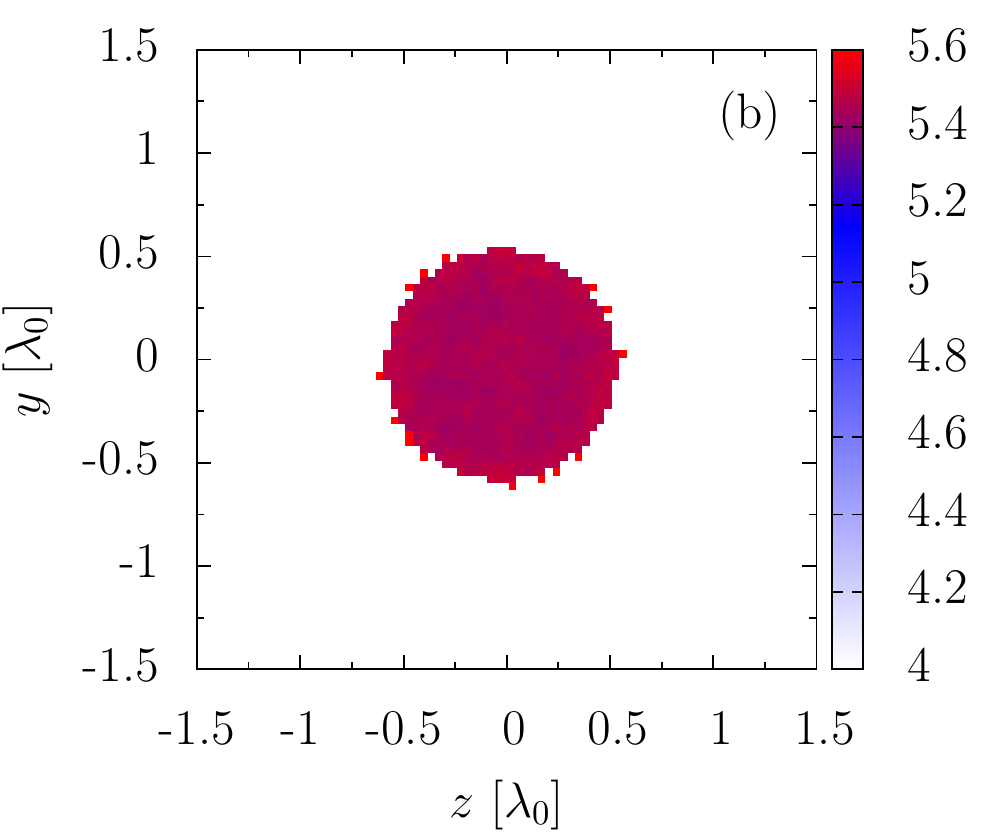}
\includegraphics[scale=0.4]{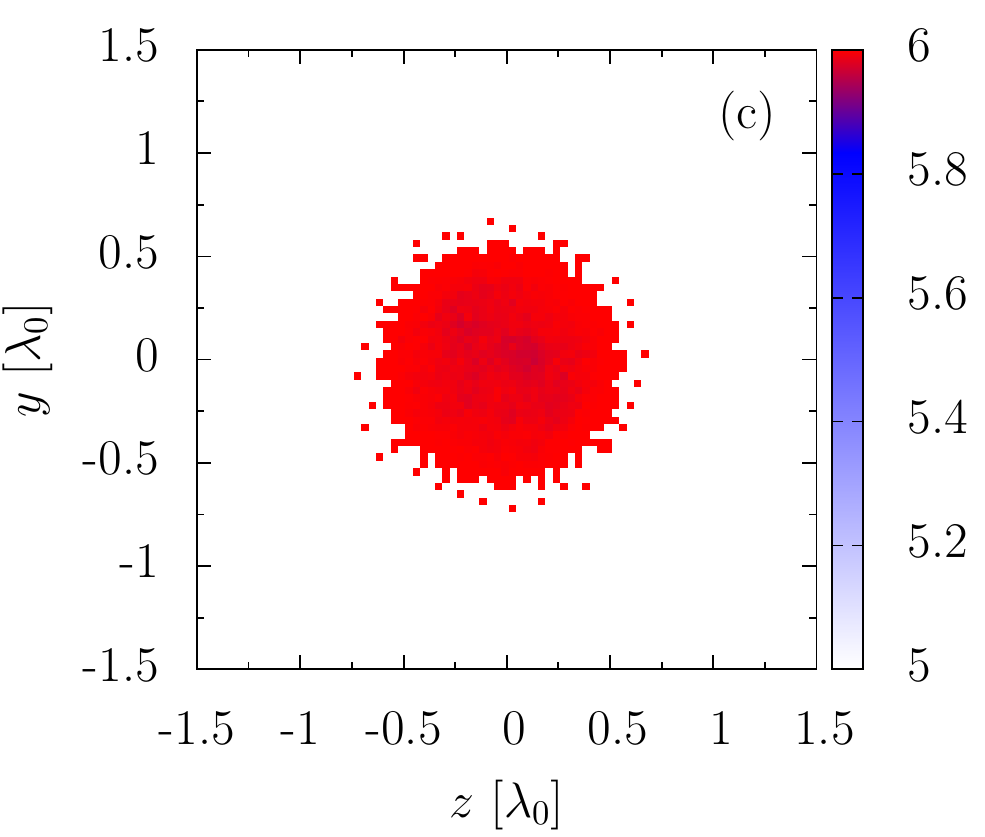}
\includegraphics[scale=0.4]{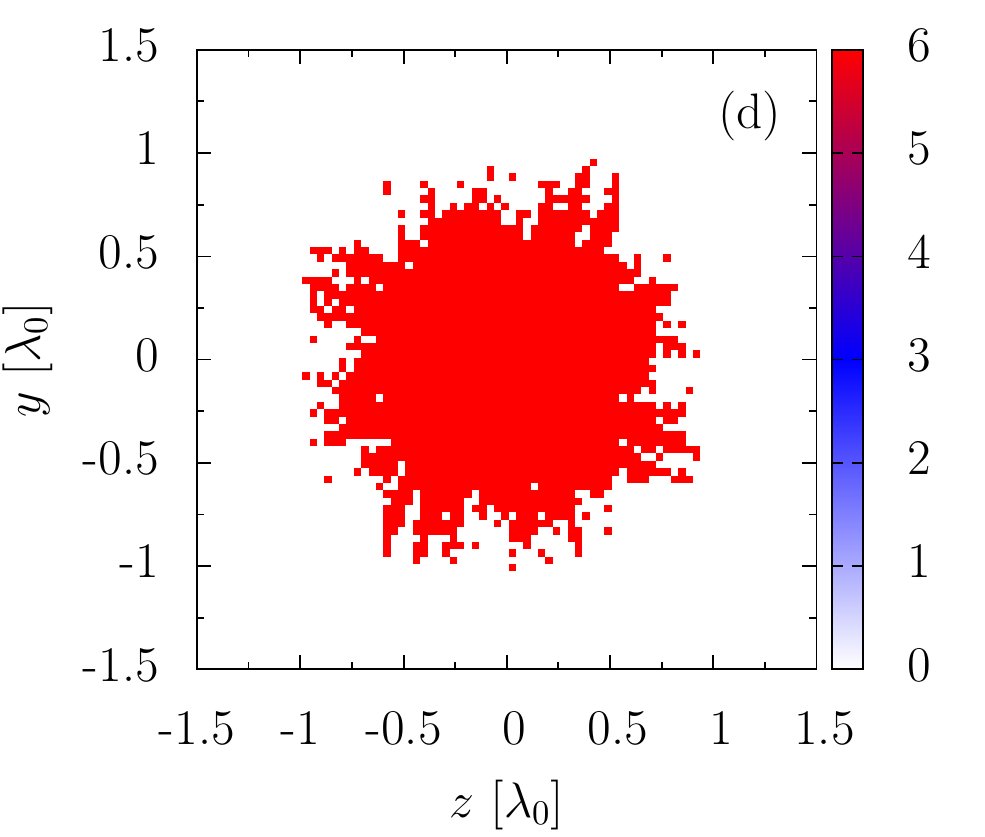}

\caption{Averaged ionization state $Z$ of the carbon ions at (a) $t = -41\ T_0$, (b) $t = -33.5\ T_0$, (c) $t = -26\ T_0$ and (d) $t = -18.5\ T_0$ before the peak of the (circularly polarized) pulse hits the surface.}
\label{a:Z_transverse_cp}
\end{figure}

This ionization structure is slightly different for a circularly polarized pulse, as we can see in figures \ref{a:Z_transverse_cp}a - \ref{a:Z_transverse_cp}d. At first, it is obvious that there is no preferred direction of the ionized surface areas (Fig. \ref{a:Z_transverse_cp}a). The surface is rather uniformly ionized over the whole circle. In the next step at $t = -33.5 T_0$ the charge states are homogenously distributed along the cross-section. While for linear polarization at $t=-26T_0$ the wire has areas with a charge of about $Z= 5.8$, Fig. \ref{a:Z_transverse_cp}c reveals a basically fully ionized wire. As before, the wire eventually expands (Fig. \ref{a:Z_transverse_cp}d).

\subsection{\label{sec:ion_void_electrons}Void electrons}

\begin{figure}[H]
\centering
  \includegraphics[scale=0.44]{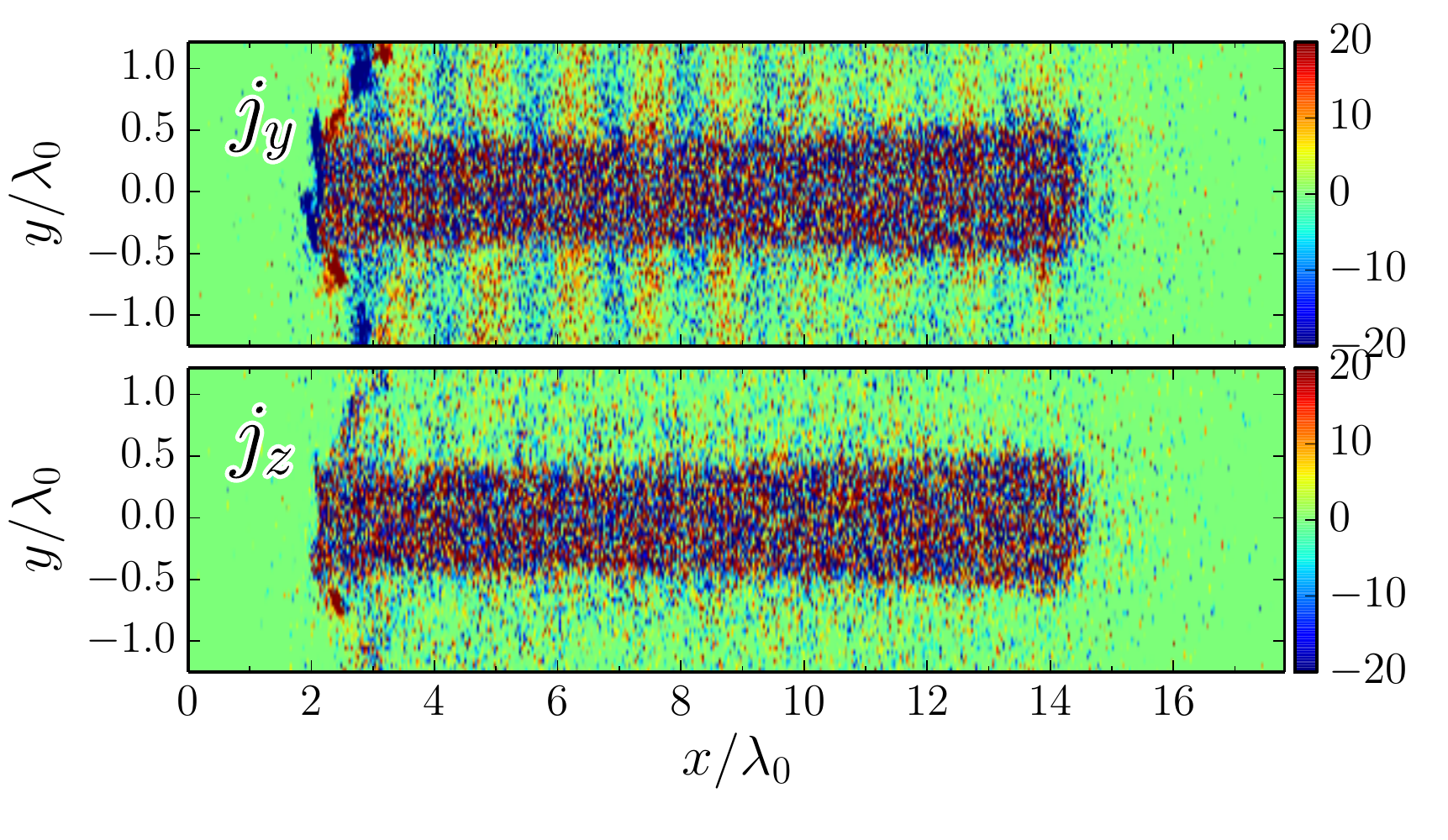}%
\caption{Longitudinal cross-sections (in the $y$-$x$ plane) of current density components $j_y$ [$e n_{cr} c$] and $j_z$ [$e n_{cr} c$] ($e$: electron charge, $n_{cr}$: critical plasma density, $c$: speed of light) at $t = 4 T_0$ after the peak of the pulse hits the surface for a linearly polarized laser}
\label{a:j_120_lp}
\end{figure}

\begin{figure}[H]
\centering
  \includegraphics[scale=0.44]{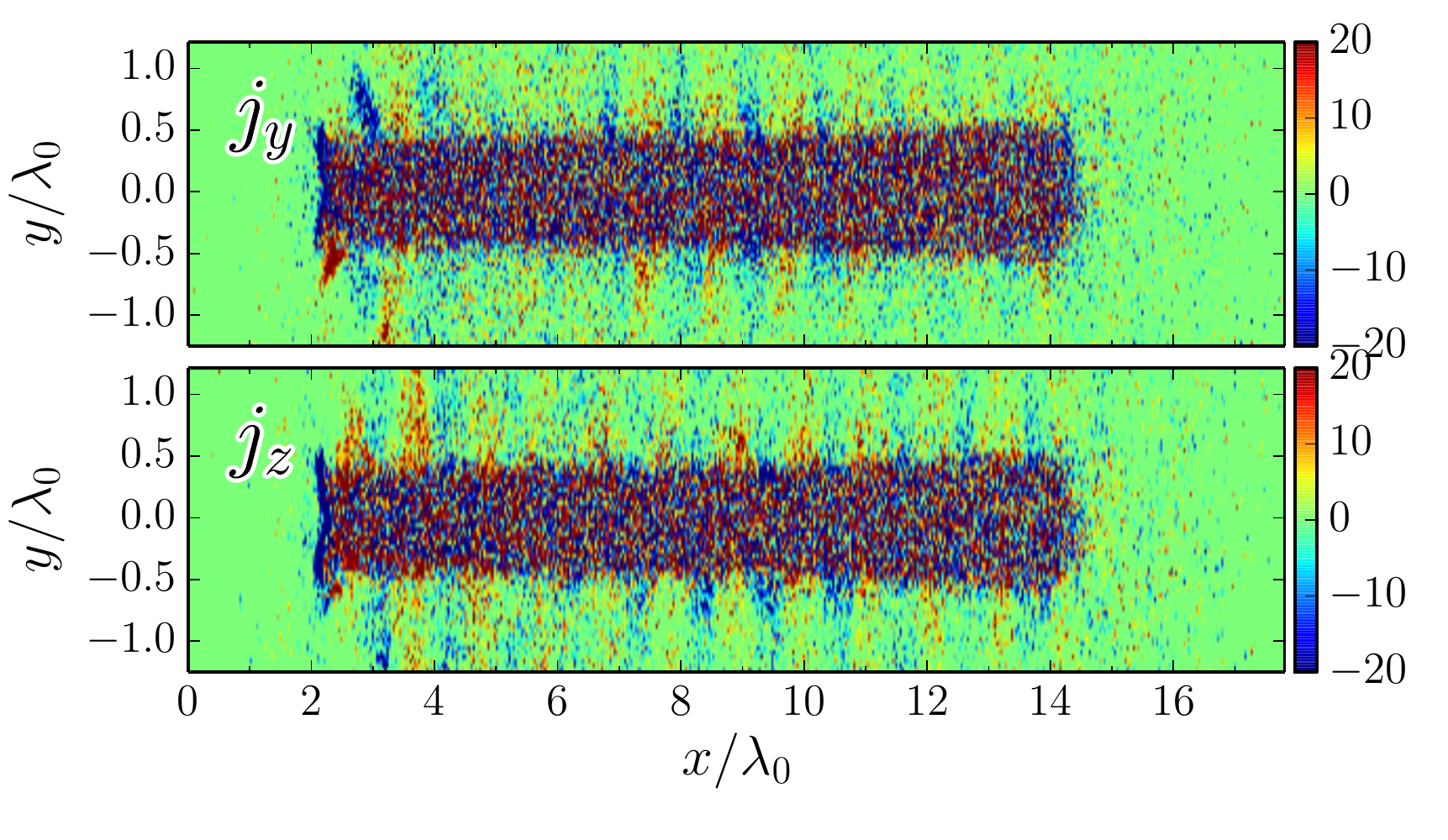}%
\caption{Longitudinal cross-sections (in the $y$-$x$ plane) of current density components $j_y$ and $j_z$ [$e n_{cr} c$] and $j_z$ [$e n_{cr} c$] ($e$: electron charge, $n_{cr}$: critical plasma density, $c$: speed of light) at $t = 4 T_0$ after the peak of the pulse hits the surface for a circularly polarized laser}
\label{a:j_120_cp}
\end{figure}

In the previous subsection we have seen that the nanowire is firstly ionized at the surface. As one could assume, the electrons that are removed from the carbon ions are pulled out of the wire into the voids. The structure of how those electrons are arranged is a trace of the electric field vector. Thus, they are ordered in the polarization plane for linear polarization. On the other hand, the released electrons are arranged on a spiral around the wire along the $x$-axis when irradiated by a circular polarized pulse. In order to see the periodicity of those electronic structure, one can employ cross-sections of the current density distributions $j_y$ and $j_z$ as show in figures \ref{a:j_120_lp} - \ref{a:j_120_cp}. The current density components $j_y$ and $j_z$ illustrate the flow of electrons in the plane perpendicular to the wire axis. Figure \ref{a:j_120_lp} confirms that there are currents in the gaps of the wires with a periodicity of one laser wavelength that move up and downwards along the polarization axis.




This becomes clear in the presentation of the $j_y$ component. However, there is no such periodic structure in the $j_z$ component. On the other hand, there are currents uniformly along both, $y$ and $z$ direction, as can bee seen by comparing the corresponding $j_y$ and $j_z$ distributions in the case of circular polarization (Fig. \ref{a:j_120_cp}). One can notice that for both polarizations these currents are restricted to the surrounding of the wire bulk since there are no visible periodic structures inside the nanowire.



\section{Conclusion}

We have investigated the ionization in laser irradiated arrays of carbon nanowires. It turns out that the strong laser pulse fully ionizes the nanowire within a time of about 30 laser cycles. We have seen how the wires are ionized along the wire axis and that a circularly polarized pulse has slightly stronger ionizing effect. We obtained that the outer layers of the nanowire are ionized first with a structure determined by the polarization. Moreover, we have shown that there are periodical transverse currents in the surrounding of the wire transporting electrons to the voids.

Further studies are required to investigate how the ionization is influenced by the wire properties as for instance the diameter, the length, the periodicity and the nanowire material. Also a variation of the laser parameters like the intensity could give more insight to the ionization process in nanowires.

\subsubsection*{Acknowledgments}
V.K. thanks John Farmer for his useful advice. This work was funded by DFG TR18, EU FP7 EUCARD-2 and by AFOSR award FA9560-14-10232.

\bibliography{quantum_electronics}

\end{document}